\PassOptionsToPackage{table}{xcolor}

\documentclass[sigconf]{acmart}
\AtBeginDocument{%
  }

\setcopyright{acmlicensed}
\copyrightyear{2025}
\acmYear{2025}
\acmDOI{XXXXXXX.XXXXXXX}
\acmConference[AHs'25]{The Augmented Humans (AHs) International Conference 2025}{March 16--20,
  2025}{Abu Dhabi, UAE}
\acmISBN{978-1-4503-XXXX-X/2018/06}

\usepackage{xcolor} % For coloring cells
\usepackage{cleveref}
\usepackage{array} % For table column width
\usepackage{makecell} % For wrapping text
\begin{document}

\title{Beyond Human: Cognitive and Physical Augmentation through AI, Robotics, and XR – Opportunities and Risks}
\author{Jie Li}
\email{jie.li@h-aisa.com}
\orcid{0000-0002-6791-104X}
\affiliation{%
  \institution{Human-AI Symbiosis Alliance}
  \city{Delft}
  \country{The Netherlands}
}

\author{Anusha Withana}
\email{anusha.withana@sydney.edu.au}
\orcid{0000-0001-6587-1278}
\affiliation{%
  \institution{University of Sydney}
  \city{Sydney}
  \country{Australia}
}

\author{Alexandra Diening}
\email{alexandra.diening@h-aisa.com}
\orcid{0009-0003-6359-2891}
\affiliation{%
  \institution{Human-AI Symbiosis Alliance}
  \city{Zurich}
  \country{Switzerland}
}

\author{Kai Kunze}
\email{kai.kunze@gmail.com}
\orcid{0000-0003-2294-3774}
\affiliation{%
  \institution{Keio University}
  \city{Tokyo}
  \country{Japan}
}

\author{Masahiko Inami}
\email{drinami@star.rcast.u-tokyo.ac.jp}
\orcid{0000-0002-8652-0730}
\affiliation{%
  \institution{University of Tokyo}
  \city{Tokyo}
  \country{Japan}
}

\renewcommand{\shortauthors}{Li et al.}

\begin{abstract}
As human augmentation technologies evolve, the convergence of AI, robotics, and extended reality (XR) is redefining human potential—enhancing cognition, perception, and physical abilities. However, these advancements also introduce ethical dilemmas, security risks, and concerns over loss of control. This workshop explores both the transformative potential and the unintended consequences of augmentation technologies. Bringing together experts from HCI, neuroscience, robotics, and ethics, we will examine real-world applications, emerging risks, and governance strategies for responsible augmentation. The session will feature keynote talks and interactive discussions, addressing topics such as AI-enhanced cognition, wearable robotics, neural interfaces, and XR-driven augmentation. By fostering multidisciplinary dialogue, this workshop aims to generate actionable insights for responsible innovation, proposing ethical frameworks to balance human empowerment with risk mitigation. We invite researchers, practitioners, and industry leaders to contribute their perspectives and help shape the future of human augmentation.
\end{abstract}



\begin{teaserfigure}
  \includegraphics[width=\textwidth]{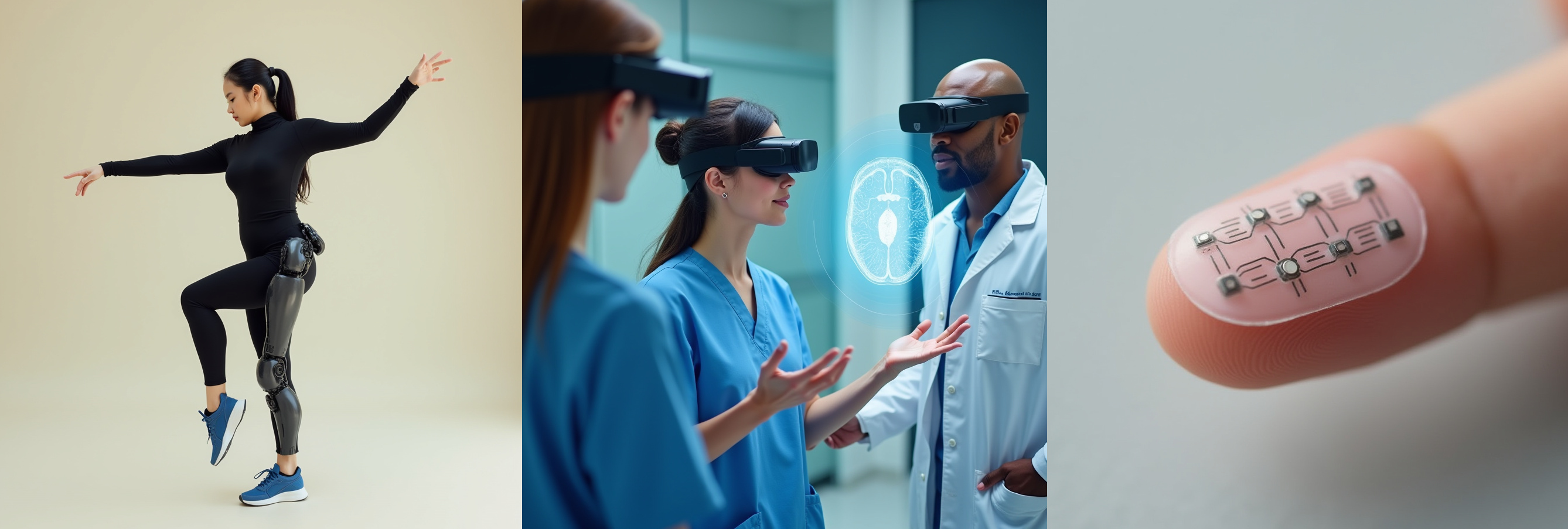}
  \caption{Possibilities of human augmentation generated by \href{https://www.freepik.com/pikaso/ai-image-generator}{Freepik AI Suite}}
  \Description{From left to right: The first image illustrates a woman with a high-tech prosthetic leg performing a graceful dance pose. She is wearing a sleek black bodysuit and blue athletic shoes, balancing elegantly on one leg. The prosthetic limb appears advanced, with a robotic design and smooth, articulated joints. The second image illustrates three medical professionals using augmented reality for brain visualization. Two healthcare workers, one male and one female, are wearing blue medical scrubs and VR headsets. A holographic image of a human brain appears between them as they engage in discussion, using Extended Reality for medical diagnostics or training. The third image illustrates a close-up of a smart electronic fingernail with embedded microcircuits. A highly detailed image of a person's fingertip showcasing a transparent artificial nail-shaped chip with tiny electronic components and circuitry, suggesting an advanced wearable technology for sensing or biometric applications.}
  \label{fig:teaser}
\end{teaserfigure}


\maketitle

\section{Introduction}

The rapid advancement of AI, robotics, and extended reality (XR) is redefining human augmentation, enhancing cognition, physical ability, perception, and human-computer collaboration. However, as augmentation systems become more autonomous and integrated into daily life, they introduce ethical, accessibility, governance, and control challenges. Key questions arise:

\begin{itemize}
\item If an AI-powered exoskeleton moves before you intend it to, is it assisting or taking control? How do we ensure augmentation enhances human agency rather than overriding it? \cite{mueller2020,diening2023}
\item If AI contributes to design or writing, who owns the final product—the human or the machine? How does AI redefine creativity and authorship? \cite{li2024_genai}
\item If only those who can afford AI-powered enhancements access them, does this innovation risk deepening inequality? How can we ensure augmentation remains inclusive? \cite{tong2024,nanayakkara2023}
\item Why do some fear AI augmentation while others embrace it? How does public perception influence adoption? \cite{stein2024attitudes}
\end{itemize}

\textbf{Human-computer symbiosis} extends beyond interaction to full integration, raising questions about bodily agency, control, and ownership \cite{mueller2020, diening2023}. Diening \cite{diening2023} emphasizes balancing automation with oversight to maintain human agency. Barbareschi et al. \cite{barbareschi2023} explore controlling multiple robotic avatars for remote work, demonstrating how shared control mechanisms allow disabled users to perform tasks remotely, but also raising concerns about agency and physical presence. Takashita et al. \cite{takashita2024} investigate \textit{Embodied Tentacle}, where users control robotic extensions like natural limbs, revealing the brain's adaptability to new augmentations. Inami et al. \cite{inami2022} propose \textit{JIZAI Body}, blending digital and physical augmentations for seamless movement, suggesting a future where robotic limbs could be interchangeable. Beyond individual augmentation, Yamamura et al. \cite{yamamura2023} introduce \textit{Social Digital Cyborgs}, which enhance teamwork, communication, and collective intelligence by using wearable robotic extensions. Zhou et al. examines how the supernumerary with different degrees of intelligence is perceived and managed by humans\cite{Zhou25}.

\textbf{Assistive technologies} are advancing accessibility in augmentation design. Tong et al. \cite{tong2024} and Withana~\cite{withana23} stress the importance of co-designing assistive tools with users, as excluding them often leads to impractical solutions. Nanayakkara et al. \cite{nanayakkara2023} propose user-driven frameworks that integrate bodily adaptation and ability constraints. Wearable robotic augmentation is also evolving, with Abadian et al. \cite{abadian2023} introducing \textit{WRLKit}, a computational tool enabling non-experts to prototype personalized robotic limbs, shifting from rigid pre-designed solutions to adaptable, user-driven augmentation.

\textbf{Generative AI (GenAI)} is reshaping creativity but introduces debates on authorship, transparency, and dependence on AI-driven workflows. Li et al. \cite{li2024_genai} examine UX professionals' perceptions of GenAI, revealing a divide between those who see it as an efficiency tool and those who fear homogenization in creative outputs. Li et al. \cite{li2024_synthetic} explore synthetic UX research, where AI-generated data simulates user feedback at scale, but this method raises concerns about reliability, bias, and the potential for replacing genuine human input with algorithmic insights. GenAI is also influencing digital embodiment. Xie et al. \cite{xie2024} demonstrate how AI-powered visualization enhances wheelchair dance choreography, expanding accessibility and artistic expression while challenging traditional notions of authorship and artistic interpretation.

\textbf{Extended Reality (XR) and haptic interfaces} redefine embodied augmentation by creating immersive, sensory-rich experiences that facilitate natural interactions. Withana et al. \cite{withana2018} introduce \textit{Tacttoo}, a thin, skin-applied haptic interface that provides real-time tactile feedback without bulky hardware. Perera et al. \cite{perera2024} develop 3D-printed electro-tactile interfaces that integrate force sensing, offering precise and intuitive haptic augmentation. Researchers are also expanding sensory augmentation beyond vision and hearing. Uyama et al. \cite{uyama2023feel} propose a system that translates music into haptic and physiological sensations, enabling multi-sensory engagement for individuals with hearing impairments.

While augmentation expands human capabilities, \textbf{governance frameworks} are needed to ensure fairness, accountability, and transparency. Cath \cite{cath2018governing} highlights regulatory oversight for public trust, and Stein et al. \cite{stein2024attitudes} examine how trust in AI is shaped by media, exposure, and personality traits. Ethical challenges such as affordability, bias, privacy, and access must be addressed to ensure augmentation benefits society rather than exacerbating divides. 

\section{Workshop Goals} 
This workshop aligns with Augmented Humans 2025 by tackling these key issues at the intersection of AI, augmentation, and ethics. Through expert talks and discussions, the workshop will:
\begin{enumerate}
\item Explore emerging augmentation technologies across cognitive, physical, and sensory domains.
\item Identify risks related to autonomy, agency, security, and bias in AI-powered augmentation.
\item Develop ethical guidelines for responsible augmentation.
\end{enumerate}

By fostering multidisciplinary dialogue, this workshop ensures that AI-driven human augmentation remains transparent, inclusive, and ethically sound.

\section{Workshop Content} 
This half-day, in-person workshop will feature a mix of keynotes, moderated discussions, and group presentations aimed at exploring the potential and risks of augmentation technologies. We plan to have the following activities:
\begin{itemize}
    \item \textbf{Keynote and Expert Talks (60 mins).} Invited speakers from AI, robotics, neuroscience, and human augmentation will present leading research findings and real-world applications. \textbf{Potential keynotes}: Prof. Masahiko Inami (Embodied Intelligence), Prof. Kai Kunze (Wearable Augmentation), Dr. Alexandra Diening (Human-AI Symbiosis).
    \item \textbf{Lightning Talks (30 mins).} Each participant will give a brief self-introduction and share their research or ethical challenges related to augmentation.
    \item \textbf{Moderated Discussion (60 mins) }. Separate participants into three groups. Each group will independently explore AI control, augmentation misuse, and regulatory needs through thought-provoking questions, including but not limited to:
    \begin{itemize}
        \item If you had full control over an additional robotic limb or avatar, what kind of tasks would you delegate to it, and what challenges might arise in seamlessly integrating it into your daily workflow?
        \item  In co-designing assistive augmentation tools, what are common mismatches between what designers assume users need and what users actually need? Can you share examples where user input drastically changed a design decision?
        \item Have you ever encountered a situation where a generative AI tool produced an outcome that was ``technically correct'' but fundamentally unusable? How did you adapt, and what does that tell us about human-AI collaboration?
        \item What is the most compelling real-world use case for haptic feedback in XR or augmentation that you've seen? What made it effective, and where do you see current limitations?
        \item  If a powerful AI-driven augmentation system could either amplify your strengths or compensate for your weaknesses, which would you choose? What ethical dilemmas arise in making such a decision?
    \end{itemize}

    \item \textbf{Group Presentation (45 min).} Each group will have 12 minutes to present their discussion results, followed by 3 minutes for Q\&A.
       
\end{itemize}

% The schedule of the workshop is presented in Table \ref{tab:schedule}.

\begin{table}[]
    \centering
    \begin{tabular}{l|l}
     \cellcolor{black}{\textcolor{white}{\textbf{Time}}}   &  \cellcolor{black}{\textcolor{white}{\textbf{Activities}}} \\
     \hline
     \textbf{09:00 – 09:30}  & Welcome \& Lighting Talks \\
     \textbf{09:30 – 10:30}  & Keynote \& Expert Talks \\
     \cellcolor{gray}\textcolor{white}{{\textbf{10:30 – 10:45}}}  & \cellcolor{gray}{\textcolor{white}{{Break}}}\\
     \textbf{10:45 – 11:45}  & Moderated Discussion \\
     \textbf{11:45 - 12:30} & Group Presentation\\
     \hline
     
    \end{tabular}
    \caption{The half-day workshop schedule}
    \label{tab:schedule}
\end{table}

\section{Expected Outcomes}
Key workshop outcomes include a \textbf{Workshop Position Paper} summarizing findings and future research directions, along with \textbf{ethical guidelines} to ensure augmentation aligns with human values. Participants will also contribute to design ideas and research challenges for further exploration. The workshop will foster collaborations between academia, industry, and policymakers to drive interdisciplinary efforts toward safe and impactful augmentation solutions. 

\section{Recruitment and Review} 
This workshop welcomes 20–25 participants, including academics, industry professionals, and researchers in HCI, AI, robotics, neuroscience, XR, and cognitive augmentation. We also invite experts in ethics, security, and policy-making who contribute to the responsible development of human augmentation technologies. We invite submissions on augmentation, including:

\begin{itemize}
    \item \textbf{Short position papers (2 pages)} on cognitive and physical augmentation.
    \item \textbf{Demonstrations} of AI-enhanced assistive technologies and augmentation prototypes.
    \item \textbf{Case studies and ethical debates} presented as position papers, posters, or slides, addressing risks, governance challenges, and societal impact.
\end{itemize}

To attract a diverse and engaged audience, we will promote the workshop through conference mailing lists, social media platforms, and direct invitations to key researchers and industry leaders in augmentation technologies. A juried process will be implemented to ensure the selection of high-quality submissions, with a focus on diversity and inclusivity in participation.

\section{Organizers Biographies}

\textbf{Jie Li} is a Human-Computer Interaction (HCI) researcher and the Chief Scientific Officer at the Human-AI Symbiosis Alliance. Her research focuses on designing experiences and developing evaluation methods for emerging technologies, including Extended Reality and Human-AI Interaction. She also writes a column for ACM Interactions, titled \textit{Bits to Bites}, where she reflects on HCI research methodologies in both academic and industry contexts.

\noindent\textbf{Anusha Withana} is an Associate Professor and an ARC DECRA fellow at the School of Computer Science, the University of Sydney. He works in the research field of human-computer interaction (HCI), mainly focusing on creating personalized enabling technologies. He is experienced in hosting workshops relating to the fabrication of new technologies.

\noindent \textbf{Alexandra Diening} is a research scientist and AI transformation expert with nearly two decades of experience leading global AI initiatives across industries, from clinical trials to education. Her work blends AI, cognitive science, and business strategy to create AI that meaningfully connects with people. She is the author of A Strategy for Human-AI Symbiosis, exploring ethical and practical AI integration. As Executive Chair of the Human-AI Symbiosis Alliance (H-AISA), she advocates for ethical AI that balances technological power with human values.

\noindent \textbf{Kai Kunze} is a professor at the Graduate School of Media Design, Keio University, specializing in wearable computing and human augmentation. His research focuses on eyewear computing, activity recognition, and amplifying human senses. He has published extensively in top venues like CHI, TOCHI, and UIST. Previously, he was an Assistant Professor at Osaka Prefecture University and conducted research at PARC, MIT Media Lab, and Sunlabs Europe.

\noindent \textbf{Masahiko Inami} is a professor at the University of Tokyo, specializing in JIZAI body editing, Augmented Humans, and entertainment engineering. He has received numerous awards, including TIME Magazine’
's ''Coolest Invention of the Year'' and MEXT's Young Scientist Award. He is a director of the Information Processing Society of Japan and the Virtual Reality Society of Japan, as well as a member of the Science Council of Japan. His latest book, Theory of JIZAI Body (Springer, 2023), explores the future of human-technology interaction.

\bibliographystyle{ACM-Reference-Format}
\bibliography{references.bib}


\begin{thebibliography}{19}

%%% the Web address.

\ifx \showCODEN    \undefined \def \showCODEN     #1{\unskip}     \fi
\ifx \showISBNx    \undefined \def \showISBNx     #1{\unskip}     \fi
\ifx \showISBNxiii \undefined \def \showISBNxiii  #1{\unskip}     \fi
\ifx \showISSN     \undefined \def \showISSN      #1{\unskip}     \fi
\ifx \showLCCN     \undefined \def \showLCCN      #1{\unskip}     \fi
\ifx \shownote     \undefined \def \shownote      #1{#1}          \fi
\ifx \showarticletitle \undefined \def \showarticletitle #1{#1}   \fi
\ifx \showURL      \undefined \def \showURL       {\relax}        \fi
% invisible to TeX
\providecommand\bibfield[2]{#2}
\providecommand\bibinfo[2]{#2}
\providecommand\natexlab[1]{#1}
\providecommand\showeprint[2][]{arXiv:#2}

\bibitem[Barbareschi et~al\mbox{.}(2023)]%
        {barbareschi2023}
\bibfield{author}{\bibinfo{person}{Giulia Barbareschi}, \bibinfo{person}{Midori
  Kawaguchi}, \bibinfo{person}{Hiroaki Kato}, \bibinfo{person}{Masato
  Nagahiro}, \bibinfo{person}{Kazuaki Takeuchi}, \bibinfo{person}{Yoshifumi
  Shiiba}, \bibinfo{person}{Shunichi Kasahara}, \bibinfo{person}{Kai Kunze},
  {and} \bibinfo{person}{Kouta Minamizawa}.} \bibinfo{year}{2023}\natexlab{}.
\newblock \showarticletitle{“I am both here and there” Parallel Control of
  Multiple Robotic Avatars by Disabled Workers in a Caf\'{e}}. In
  \bibinfo{booktitle}{\emph{Proceedings of the 2023 CHI Conference on Human
  Factors in Computing Systems}} (Hamburg, Germany) \emph{(\bibinfo{series}{CHI
  '23})}. \bibinfo{publisher}{Association for Computing Machinery},
  \bibinfo{address}{New York, NY, USA}, Article \bibinfo{articleno}{75},
  \bibinfo{numpages}{17}~pages.
\newblock
\showISBNx{9781450394215}
\href{https://doi.org/10.1145/3544548.3581124}{doi:\nolinkurl{10.1145/3544548.3581124}}


\bibitem[Cath(2018)]%
        {cath2018governing}
\bibfield{author}{\bibinfo{person}{Corinne Cath}.}
  \bibinfo{year}{2018}\natexlab{}.
\newblock \showarticletitle{Governing artificial intelligence: ethical, legal
  and technical opportunities and challenges}.
\newblock \bibinfo{journal}{\emph{Philosophical Transactions of the Royal
  Society A: Mathematical, Physical and Engineering Sciences}}
  \bibinfo{volume}{376}, \bibinfo{number}{2133} (\bibinfo{year}{2018}),
  \bibinfo{pages}{20180080}.
\newblock


\bibitem[Dieing(2024)]%
        {diening2023}
\bibfield{author}{\bibinfo{person}{Alexandra Dieing}.}
  \bibinfo{year}{2024}\natexlab{}.
\newblock \bibinfo{booktitle}{\emph{A Strategy for Human-AI Symbiosis.:
  Concepts, Tools, and Business Models for the New AI Game}}.
\newblock \bibinfo{publisher}{Independently published},
  \bibinfo{address}{Zurich Switzerland}.
\newblock
\showISBNx{979-8345430873}


\bibitem[Inami et~al\mbox{.}(2022)]%
        {inami2022}
\bibfield{author}{\bibinfo{person}{Masahiko Inami}, \bibinfo{person}{Daisuke
  Uriu}, \bibinfo{person}{Zendai Kashino}, \bibinfo{person}{Shigeo Yoshida},
  \bibinfo{person}{Hiroto Saito}, \bibinfo{person}{Azumi Maekawa}, {and}
  \bibinfo{person}{Michiteru Kitazaki}.} \bibinfo{year}{2022}\natexlab{}.
\newblock \showarticletitle{Cyborgs, Human Augmentation, Cybernetics, and JIZAI
  Body}. In \bibinfo{booktitle}{\emph{Proceedings of the Augmented Humans
  International Conference 2022}} (Kashiwa, Chiba, Japan)
  \emph{(\bibinfo{series}{AHs '22})}. \bibinfo{publisher}{Association for
  Computing Machinery}, \bibinfo{address}{New York, NY, USA},
  \bibinfo{pages}{230–242}.
\newblock
\showISBNx{9781450396325}
\href{https://doi.org/10.1145/3519391.3519401}{doi:\nolinkurl{10.1145/3519391.3519401}}


\bibitem[Li(2024)]%
        {li2024_synthetic}
\bibfield{author}{\bibinfo{person}{Jie Li}.} \bibinfo{year}{2024}\natexlab{}.
\newblock \showarticletitle{How Far Can We Go with Synthetic User Experience
  Research?}
\newblock \bibinfo{journal}{\emph{Interactions}} \bibinfo{volume}{31},
  \bibinfo{number}{3} (\bibinfo{date}{May} \bibinfo{year}{2024}),
  \bibinfo{pages}{26–29}.
\newblock
\showISSN{1072-5520}
\href{https://doi.org/10.1145/3653682}{doi:\nolinkurl{10.1145/3653682}}


\bibitem[Li et~al\mbox{.}(2024)]%
        {li2024_genai}
\bibfield{author}{\bibinfo{person}{Jie Li}, \bibinfo{person}{Hancheng Cao},
  \bibinfo{person}{Laura Lin}, \bibinfo{person}{Youyang Hou},
  \bibinfo{person}{Ruihao Zhu}, {and} \bibinfo{person}{Abdallah El~Ali}.}
  \bibinfo{year}{2024}\natexlab{}.
\newblock \showarticletitle{User Experience Design Professionals’ Perceptions
  of Generative Artificial Intelligence}. In
  \bibinfo{booktitle}{\emph{Proceedings of the 2024 CHI Conference on Human
  Factors in Computing Systems}} (Honolulu, HI, USA)
  \emph{(\bibinfo{series}{CHI '24})}. \bibinfo{publisher}{Association for
  Computing Machinery}, \bibinfo{address}{New York, NY, USA}, Article
  \bibinfo{articleno}{381}, \bibinfo{numpages}{18}~pages.
\newblock
\showISBNx{9798400703300}
\href{https://doi.org/10.1145/3613904.3642114}{doi:\nolinkurl{10.1145/3613904.3642114}}


\bibitem[Mueller et~al\mbox{.}(2020)]%
        {mueller2020}
\bibfield{author}{\bibinfo{person}{Florian~Floyd Mueller},
  \bibinfo{person}{Pedro Lopes}, \bibinfo{person}{Paul Strohmeier},
  \bibinfo{person}{Wendy Ju}, \bibinfo{person}{Caitlyn Seim},
  \bibinfo{person}{Martin Weigel}, \bibinfo{person}{Suranga Nanayakkara},
  \bibinfo{person}{Marianna Obrist}, \bibinfo{person}{Zhuying Li},
  \bibinfo{person}{Joseph Delfa}, \bibinfo{person}{Jun Nishida},
  \bibinfo{person}{Elizabeth~M. Gerber}, \bibinfo{person}{Dag Svanaes},
  \bibinfo{person}{Jonathan Grudin}, \bibinfo{person}{Stefan Greuter},
  \bibinfo{person}{Kai Kunze}, \bibinfo{person}{Thomas Erickson},
  \bibinfo{person}{Steven Greenspan}, \bibinfo{person}{Masahiko Inami},
  \bibinfo{person}{Joe Marshall}, \bibinfo{person}{Harald Reiterer},
  \bibinfo{person}{Katrin Wolf}, \bibinfo{person}{Jochen Meyer},
  \bibinfo{person}{Thecla Schiphorst}, \bibinfo{person}{Dakuo Wang}, {and}
  \bibinfo{person}{Pattie Maes}.} \bibinfo{year}{2020}\natexlab{}.
\newblock \showarticletitle{Next Steps for Human-Computer Integration}. In
  \bibinfo{booktitle}{\emph{Proceedings of the 2020 CHI Conference on Human
  Factors in Computing Systems}} (Honolulu, HI, USA)
  \emph{(\bibinfo{series}{CHI '20})}. \bibinfo{publisher}{Association for
  Computing Machinery}, \bibinfo{address}{New York, NY, USA},
  \bibinfo{pages}{1–15}.
\newblock
\showISBNx{9781450367080}
\href{https://doi.org/10.1145/3313831.3376242}{doi:\nolinkurl{10.1145/3313831.3376242}}


\bibitem[Nanayakkara et~al\mbox{.}(2023)]%
        {nanayakkara2023}
\bibfield{author}{\bibinfo{person}{Suranga~Chandima Nanayakkara},
  \bibinfo{person}{Masahiko Inami}, \bibinfo{person}{Florian Mueller},
  \bibinfo{person}{Jochen Huber}, \bibinfo{person}{Chitralekha Gupta},
  \bibinfo{person}{Christophe Jouffrais}, \bibinfo{person}{Kai Kunze},
  \bibinfo{person}{Rakesh Patibanda}, \bibinfo{person}{Samantha W~T Chan},
  {and} \bibinfo{person}{Moritz~Alexander Messerschmidt}.}
  \bibinfo{year}{2023}\natexlab{}.
\newblock \showarticletitle{Exploring the Design Space of Assistive
  Augmentation}. In \bibinfo{booktitle}{\emph{Proceedings of the Augmented
  Humans International Conference 2023}} (Glasgow, United Kingdom)
  \emph{(\bibinfo{series}{AHs '23})}. \bibinfo{publisher}{Association for
  Computing Machinery}, \bibinfo{address}{New York, NY, USA},
  \bibinfo{pages}{371–373}.
\newblock
\showISBNx{9781450399845}
\href{https://doi.org/10.1145/3582700.3582729}{doi:\nolinkurl{10.1145/3582700.3582729}}


\bibitem[Perera et~al\mbox{.}(2024)]%
        {perera2024}
\bibfield{author}{\bibinfo{person}{Praneeth~Bimsara Perera},
  \bibinfo{person}{Hansa Marasinghe}, \bibinfo{person}{Taiki Takami},
  \bibinfo{person}{Hiroyuki Kajimoto}, {and} \bibinfo{person}{Anusha Withana}.}
  \bibinfo{year}{2024}\natexlab{}.
\newblock \showarticletitle{Integrating Force Sensing with Electro-Tactile
  Feedback in 3D Printed Haptic Interfaces}. In
  \bibinfo{booktitle}{\emph{Proceedings of the 2024 ACM International Symposium
  on Wearable Computers}} (Melbourne VIC, Australia)
  \emph{(\bibinfo{series}{ISWC '24})}. \bibinfo{publisher}{Association for
  Computing Machinery}, \bibinfo{address}{New York, NY, USA},
  \bibinfo{pages}{48–54}.
\newblock
\showISBNx{9798400710599}
\href{https://doi.org/10.1145/3675095.3676612}{doi:\nolinkurl{10.1145/3675095.3676612}}


\bibitem[Saberpour~Abadian et~al\mbox{.}(2023)]%
        {abadian2023}
\bibfield{author}{\bibinfo{person}{Artin Saberpour~Abadian},
  \bibinfo{person}{Ata Otaran}, \bibinfo{person}{Martin Schmitz},
  \bibinfo{person}{Marie Muehlhaus}, \bibinfo{person}{Rishabh Dabral},
  \bibinfo{person}{Diogo Luvizon}, \bibinfo{person}{Azumi Maekawa},
  \bibinfo{person}{Masahiko Inami}, \bibinfo{person}{Christian Theobalt}, {and}
  \bibinfo{person}{J\"{u}rgen Steimle}.} \bibinfo{year}{2023}\natexlab{}.
\newblock \showarticletitle{Computational Design of Personalized Wearable
  Robotic Limbs}. In \bibinfo{booktitle}{\emph{Proceedings of the 36th Annual
  ACM Symposium on User Interface Software and Technology}} (San Francisco, CA,
  USA) \emph{(\bibinfo{series}{UIST '23})}. \bibinfo{publisher}{Association for
  Computing Machinery}, \bibinfo{address}{New York, NY, USA}, Article
  \bibinfo{articleno}{68}, \bibinfo{numpages}{13}~pages.
\newblock
\showISBNx{9798400701320}
\href{https://doi.org/10.1145/3586183.3606748}{doi:\nolinkurl{10.1145/3586183.3606748}}


\bibitem[Stein et~al\mbox{.}(2024)]%
        {stein2024attitudes}
\bibfield{author}{\bibinfo{person}{Jan-Philipp Stein}, \bibinfo{person}{Tanja
  Messingschlager}, \bibinfo{person}{Timo Gnambs}, \bibinfo{person}{Fabian
  Hutmacher}, {and} \bibinfo{person}{Markus Appel}.}
  \bibinfo{year}{2024}\natexlab{}.
\newblock \showarticletitle{Attitudes towards AI: measurement and associations
  with personality}.
\newblock \bibinfo{journal}{\emph{Scientific Reports}} \bibinfo{volume}{14},
  \bibinfo{number}{1} (\bibinfo{year}{2024}), \bibinfo{pages}{2909}.
\newblock


\bibitem[Takashita et~al\mbox{.}(2024)]%
        {takashita2024}
\bibfield{author}{\bibinfo{person}{Shuto Takashita}, \bibinfo{person}{Ken
  Arai}, \bibinfo{person}{Hiroto Saito}, \bibinfo{person}{Michiteru Kitazaki},
  {and} \bibinfo{person}{Masahiko Inami}.} \bibinfo{year}{2024}\natexlab{}.
\newblock \showarticletitle{Embodied Tentacle: Mapping Design to Control of
  Non-Analogous Body Parts with the Human Body}. In
  \bibinfo{booktitle}{\emph{Proceedings of the 2024 CHI Conference on Human
  Factors in Computing Systems}} (Honolulu, HI, USA)
  \emph{(\bibinfo{series}{CHI '24})}. \bibinfo{publisher}{Association for
  Computing Machinery}, \bibinfo{address}{New York, NY, USA}, Article
  \bibinfo{articleno}{222}, \bibinfo{numpages}{19}~pages.
\newblock
\showISBNx{9798400703300}
\href{https://doi.org/10.1145/3613904.3642340}{doi:\nolinkurl{10.1145/3613904.3642340}}


\bibitem[Tong et~al\mbox{.}(2024)]%
        {tong2024}
\bibfield{author}{\bibinfo{person}{Adele Tong}, \bibinfo{person}{Zhanna
  Sarsenbayeva}, \bibinfo{person}{Jorge Goncalves}, \bibinfo{person}{Hideki
  Koike}, \bibinfo{person}{Masahiko Inami}, \bibinfo{person}{Alistair McEwan},
  {and} \bibinfo{person}{Anusha Withana}.} \bibinfo{year}{2024}\natexlab{}.
\newblock \showarticletitle{Developing Strategies for Co-designing Assistive
  Augmentation Technologies}. In \bibinfo{booktitle}{\emph{Proceedings of the
  Augmented Humans International Conference 2024}} (Melbourne, VIC, Australia)
  \emph{(\bibinfo{series}{AHs '24})}. \bibinfo{publisher}{Association for
  Computing Machinery}, \bibinfo{address}{New York, NY, USA},
  \bibinfo{pages}{324–326}.
\newblock
\showISBNx{9798400709807}
\href{https://doi.org/10.1145/3652920.3653038}{doi:\nolinkurl{10.1145/3652920.3653038}}


\bibitem[Uyama et~al\mbox{.}(2023)]%
        {uyama2023feel}
\bibfield{author}{\bibinfo{person}{Aoi Uyama}, \bibinfo{person}{Danny Hynds},
  \bibinfo{person}{Dingding Zheng}, \bibinfo{person}{George Chernyshov},
  \bibinfo{person}{Tatsuya Saito}, \bibinfo{person}{Kai Kunze}, {and}
  \bibinfo{person}{Kouta Minamizawa}.} \bibinfo{year}{2023}\natexlab{}.
\newblock \showarticletitle{Feel What You Don’t Hear: A New Framework for
  Non-Aural Music Experiences}. In \bibinfo{booktitle}{\emph{Proceedings of the
  International Conference on New Interfaces for Musical Expression}}.
  \bibinfo{publisher}{Association for Computing Machinery},
  \bibinfo{address}{New York, NY, USA}, \bibinfo{pages}{560--565}.
\newblock


\bibitem[Withana(2023)]%
        {withana23}
\bibfield{author}{\bibinfo{person}{Anusha Withana}.}
  \bibinfo{year}{2023}\natexlab{}.
\newblock \showarticletitle{Co-Designing Personalized Assistive Devices Using
  Personal Fabrication}.
\newblock \bibinfo{journal}{\emph{Commun. ACM}} \bibinfo{volume}{66},
  \bibinfo{number}{7} (\bibinfo{date}{jun} \bibinfo{year}{2023}),
  \bibinfo{pages}{89–90}.
\newblock
\showISSN{0001-0782}
\href{https://doi.org/10.1145/3589150}{doi:\nolinkurl{10.1145/3589150}}


\bibitem[Withana et~al\mbox{.}(2018)]%
        {withana2018}
\bibfield{author}{\bibinfo{person}{Anusha Withana}, \bibinfo{person}{Daniel
  Groeger}, {and} \bibinfo{person}{J\"{u}rgen Steimle}.}
  \bibinfo{year}{2018}\natexlab{}.
\newblock \showarticletitle{Tacttoo: A Thin and Feel-Through Tattoo for On-Skin
  Tactile Output}. In \bibinfo{booktitle}{\emph{Proceedings of the 31st Annual
  ACM Symposium on User Interface Software and Technology}} (Berlin, Germany)
  \emph{(\bibinfo{series}{UIST '18})}. \bibinfo{publisher}{Association for
  Computing Machinery}, \bibinfo{address}{New York, NY, USA},
  \bibinfo{pages}{365–378}.
\newblock
\showISBNx{9781450359481}
\href{https://doi.org/10.1145/3242587.3242645}{doi:\nolinkurl{10.1145/3242587.3242645}}


\bibitem[Xie et~al\mbox{.}(2024)]%
        {xie2024}
\bibfield{author}{\bibinfo{person}{Yurui Xie}, \bibinfo{person}{Giulia
  Barbareschi}, \bibinfo{person}{Kai Kunze}, {and} \bibinfo{person}{Masa
  Inakage}.} \bibinfo{year}{2024}\natexlab{}.
\newblock \showarticletitle{Exploring Digital Embodiment in Wheelchair Dance
  with Generative AI}. In \bibinfo{booktitle}{\emph{Proceedings of the 13th
  International Conference on the Internet of Things}} (Nagoya, Japan)
  \emph{(\bibinfo{series}{IoT '23})}. \bibinfo{publisher}{Association for
  Computing Machinery}, \bibinfo{address}{New York, NY, USA},
  \bibinfo{pages}{224–227}.
\newblock
\showISBNx{9798400708541}
\href{https://doi.org/10.1145/3627050.3631576}{doi:\nolinkurl{10.1145/3627050.3631576}}


\bibitem[Yamamura et~al\mbox{.}(2023)]%
        {yamamura2023}
\bibfield{author}{\bibinfo{person}{Nahoko Yamamura}, \bibinfo{person}{Daisuke
  Uriu}, \bibinfo{person}{Mitsuru Muramatsu}, \bibinfo{person}{Yusuke
  Kamiyama}, \bibinfo{person}{Zendai Kashino}, \bibinfo{person}{Shin Sakamoto},
  \bibinfo{person}{Naoki Tanaka}, \bibinfo{person}{Toma Tanigawa},
  \bibinfo{person}{Akiyoshi Onishi}, \bibinfo{person}{Shigeo Yoshida},
  \bibinfo{person}{Shunji Yamanaka}, {and} \bibinfo{person}{Masahiko Inami}.}
  \bibinfo{year}{2023}\natexlab{}.
\newblock \showarticletitle{Social Digital Cyborgs: The Collaborative Design
  Process of JIZAI ARMS}. In \bibinfo{booktitle}{\emph{Proceedings of the 2023
  CHI Conference on Human Factors in Computing Systems}} (Hamburg, Germany)
  \emph{(\bibinfo{series}{CHI '23})}. \bibinfo{publisher}{Association for
  Computing Machinery}, \bibinfo{address}{New York, NY, USA}, Article
  \bibinfo{articleno}{369}, \bibinfo{numpages}{19}~pages.
\newblock
\showISBNx{9781450394215}
\href{https://doi.org/10.1145/3544548.3581169}{doi:\nolinkurl{10.1145/3544548.3581169}}


\bibitem[Zhou et~al\mbox{.}(2025)]%
        {Zhou25}
\bibfield{author}{\bibinfo{person}{Hongyu Zhou}, \bibinfo{person}{Tom Kip},
  \bibinfo{person}{Andrea Bianchi}, \bibinfo{person}{Zhanna Sarsenbayeva},
  {and} \bibinfo{person}{Anusha Withana}.} \bibinfo{year}{2025}\natexlab{}.
\newblock \showarticletitle{Juggling Extra Limbs: Identifying Control
  Strategies for Supernumerary Multi-Arms in Virtual Reality}. In
  \bibinfo{booktitle}{\emph{Proceedings of the 2025 CHI Conference on Human
  Factors in Computing Systems}} \emph{(\bibinfo{series}{CHI '25})}.
  \bibinfo{publisher}{Association for Computing Machinery},
  \bibinfo{address}{New York, NY, USA}.
\newblock
\showISBNx{9781450394215}
\href{https://doi.org/10.1145/3706598.3713647}{doi:\nolinkurl{10.1145/3706598.3713647}}


\end{thebibliography}


\end{document}